\documentclass[12pt]{amsart}

\usepackage{amsmath,amssymb, bm}
\usepackage{graphicx}
\usepackage[linesnumbered,lined,boxed,commentsnumbered, ruled]{algorithm2e}

\usepackage[table]{xcolor}

\usepackage{kbordermatrix}

\usepackage[english]{babel}
\usepackage[hidelinks]{hyperref}
\newcommand{\be}{\begin{equation}}
\newcommand{\ee}{\end{equation}}
\newcommand{\mk}{\langle k\rangle}
\newcommand{\inn}{\text{in}}
\newcommand{\out}{\text{out}}
\newcommand{\mean}[1]{\left< #1 \right>}
\newcommand{\crossed}{{\backslash \hspace{-4.33pt}/}}

\begin{document}
\title{Degree assortativity in networks of spiking neurons}% Force line breaks with \\

\author{Christian Bl{\"a}sche}
\email{c.blasche@massey.ac.nz}
\address{ School of Natural and Computational Sciences,
Massey University, Private Bag 102-904 NSMC, Auckland, New Zealand.
 }

\author{Shawn Means}
\email{S.Means@massey.ac.nz}
\address{ School of Natural and Computational Sciences,
Massey University, Private Bag 102-904 NSMC, Auckland, New Zealand.
 }

\author{Carlo R. Laing}
\email{c.r.laing@massey.ac.nz}
\address{ School of Natural and Computational Sciences,
Massey University, Private Bag 102-904 NSMC, Auckland, New Zealand. \\
phone: +64-9-414 0800 extn. 43512
fax: +64-9-4418136 }

\date{\today}
%\pacs{05.45.Xt, 05.45.Ac}

\begin{abstract}
Degree assortativity refers to the increased or decreased probability of connecting two
neurons based on their in- or out-degrees, relative to what would be expected by chance. 
We investigate the effects of such assortativity in a network of theta neurons. The Ott/Antonsen
ansatz is used to derive equations for the expected state of each neuron, and these equations
are then coarse-grained in degree space. We generate families of effective connectivity matrices
parametrised by assortativity coefficient and use SVD decompositions of
these to efficiently perform numerical bifurcation
analysis of the coarse-grained equations. We find that of the four possible types of degree
assortativity, two have no effect on the networks' dynamics, while the other two can have 
a significant effect.

\end{abstract}

\maketitle

%\maketitle

\section{Introduction}
Our nervous system consists of a vast network of interconnected neurons.
The network structure is dynamic and connections are formed or removed according to their usage.
% to serve specific functions of that region.
%The effects of the structure of a network of neurons on its dynamics is a large 
%topic of current interest~\cite{vashou12,lamsmi10,marhou16,rox11,schkih15,setdeg17,litdoi12}.
Much effort has been put into creating a map of all neuronal interconnections; a so-called brain atlas or \textit{connectome}. Given such a network
there are many structural features and measures that one can use to characterise it, e.g. betweenness, centrality, average path-length and clustering coefficient~\cite{new03}.

Obtaining these measures in actual physiological systems is challenging to say the least; nevertheless, insights into intrinsic connectivity preferences of neurons were observed via their growth in culture~\cite{desen14,teller2014}. Neurons with similar numbers of processes (e.g., synapses and dendrites) tend to establish links with each other -- akin to socialites associating in groups and vice-versa. Such an assortativity, typically referred to as a positive assortativity, or a tendency of elements with similar properties to mutually connect, emerges as a strong preference throughout the stages of the cultured neuron development. Furthermore, this preferential attachment between highly-connected neurons is suggested to fortify the neuronal network against disruption or damage~\cite{teller2014}. Moreover, a similar positive assortativity is inferred in human central nervous systems as well~\cite{desen14} at both a structural and functional level, where a central ``core'' in the human cerebral cortex may be the basis for shaping overall brain dynamics~\cite{hagmann2008}. It seems that in no instance, however, is the directional flow of information (e.g., from upstream neuron via axon to synapse and downstream neuron) observed -- either in culture or \emph{in situ}.

Little is known about if and how this positive assortativity or its negative analogue (a tendency for highly connected elements to link with sparsely connected -- akin to a ``hub-and-spoke'' system) influences the dynamics of the neurons networked together. We here consider these effects of assortativity with a network of idealised mathematical neurons known as theta neurons, and represent their connections of downstream (outbound) and upstream (inbound) synapses in a directed graph of nodes and edges. Each neuron then is a node with in- and out-degrees depicting the number of such connections within the network.
%Little is known though about if and how they affect the network dynamics.
%Here we consider the effects of degree assortativity in a large network of theta neurons and aim to get a qualitatively understanding of its influence on neuronal activity.
Assortativity in this context refers to the probability that a neuron with a given
in- and out-degree connects to another neuron with a given in- and out-degree. If this probability
is what one would expect by chance, given the neurons' degrees (and this is the case for all
pairs of neurons),
the network is referred to as neutrally assortative.
If the probability is higher (lower) than one would expect by chance --- for all pairs ---
the network
is assortative (disassortative).
Interchangeably, we will use the term postive assortativity (negative assortativity).

\begin{figure}
\includegraphics[width=.7\linewidth]{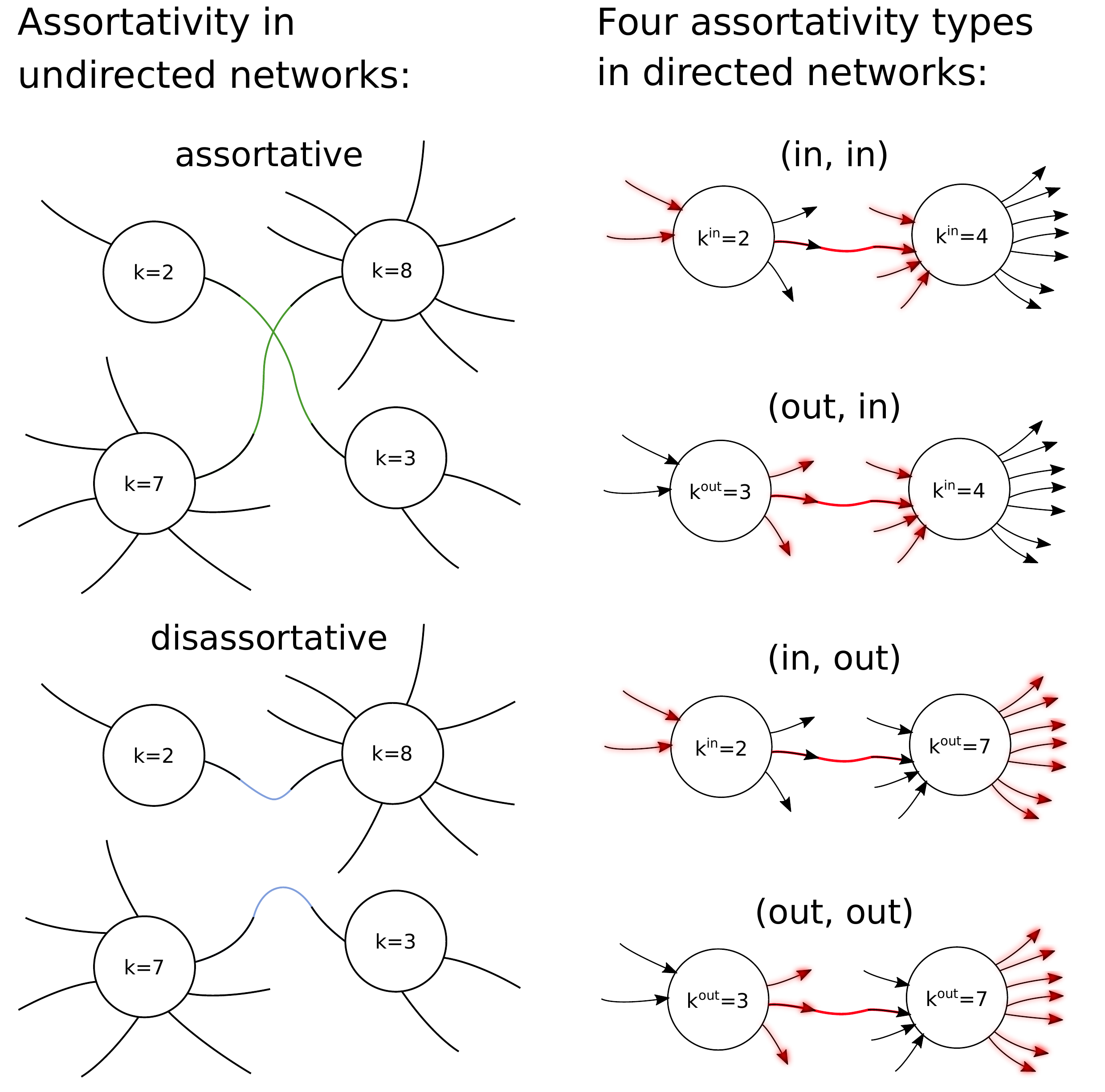}
\caption{Assortativity in undirected and directed networks. An undirected network (left column)
is assortative if high degree nodes are more likely to be connected to high degree nodes,
and low to low, than by chance (top left). Such a network is disassortative if the opposite
occurs (bottom left).
In directed networks (right column)
 there are four possible kinds of assortativity. 
The probability of a connection (red) is thus influenced by the number of red shaded links of the sending (left) and receiving (right) node.}
\label{fig_assort_type_schematic1}
\end{figure}

Assortativity has often been studied in undirected networks, where a node simply has a
degree, rather than in- and out-degrees (the number of connections to and from a node,
respectively)~\cite{resott14,new03,new02}. 
Since neurons form {\em directed} connections, there are four types of assortativity to
consider~\cite{Fosfos10}: between either the in- or out-degree of a presynaptic neuron, and
either the in- or out-degree of a postsynaptic neuron (Figure~\ref{fig_assort_type_schematic1}). 

We are aware of only a small number
of previous
studies in this area~\cite{schkih15,kahsok17,avalos2012,defranciscis2011}.
K{\"a}hne et al. \cite{kahsok17} considered networks with equal in- and out-degrees
and investigated
degree assortativity, 
effectively correlating both in- and out-degrees of pre- and post-synaptic
neurons. They mostly considered networks with discrete time and a Heaviside firing rate, 
i.e.~a McCulloch-Pitts model~\cite{mccpit48}. They found that
positive assortativity created new fixed points of the model dynamics.
Schmeltzer et al.~\cite{schkih15} also consider networks with 
equal in- and out-degrees and investigated
degree assortativity. These authors considered leaky integrate-and-fire 
neurons and derived approximate self-consistency
equations governing the steady state neuron firing rates. They found, among other
things, that positive assortativity increased the firing rates of high-degree neurons and decreased
that of low-degree ones. Positive assortativity also seemed to make the network more capable
of sustained activity when the external input to the network was low. De Fransciscis 
et al.~\cite{defranciscis2011} considered assortative mixing of a field of binary neurons, or Hopfield networks. They concluded that assortativity of such simple model neurons exhibited associative memory (similar to bit fields of a magnetic storage medium), and robustly so in the presence of noise that negatively assortative networks failed to withstand. Avalos-Gaytan et al.~\cite{avalos2012} considered the effects of dynamic weightings between Kuramoto oscillators --- effectively a dynamically evolving network --- on assortativity. They observed that if the strength of connections between
oscillators increased when they were synchronised, a strong positive assortativity evolved in the network, suggesting a potential mechanism for the creation of assortative networks, as observed in cultured neurons mentioned above, and as we study here.

To briefly summarise our results, we find that only two out of the four types
of degree assorativity have any influence on the network's dynamics: those when the in-degree
of a presynaptic neuron is correlated with either the in- or out-degree of a postsynaptic neuron.
Of these two, (in,in)-assortativity has a greater effect than (in,out)-assortativity.
For both cases, negative assortativity widens the parameter range for which the network
is bistable (for excitatory coupling) or undergoes oscillations in mean firing rate
(for inhibitory coupling), and positive assortativity has the opposite effect. 

Our work is similar in some respects to that of Restrepo and Ott~\cite{resott14} who considered
degree assortativity in a network of Kuramoto-type phase oscillators. They found that for
positive assortativity, as the strength of connections between oscillators was increased the
network could undergo bifurcations leading to oscillations in the order parameter,
in contrast to the usual scenario that occurs for no assortativity. However, their network
was {\em undirected}, and thus there is only one type of degree assortativity possible.

%Experimental evidence for positive degree assortativity includes~\cite{desen14},
%who examined a neuronal culture but did not determine
%directionality of connections. Egu\'{\i}luz et al.~\cite{eguchi05} also found
%evidence of positive assortativity 
%in the brain, and did not determine directionality of connections. They also
%found evidence for a power law degree distribution, as we consider here. 
%In summary, 
%we find that only two out of four degree assortativity cases influence the dynamics, i.e. when the in-degree of the sending neuron is correlated to either in- or out-degree of the receiving neuron.
%Among those two, (in,in)-assortativity pushes bifurcations way stronger and in both scenarios of either purely excitatory or inhibitory coupling negative assortativity is widening the bistable area or the region of oscillations respectively, whereas positive assortativity has the opposite effect.

%\cite{pirpro12}

The outline of the paper is as follows. In Sec.~\ref{sec:model} we present the model
and then derive several approximate descriptions of its dynamics. In Sec.~\ref{sec:assort}
we describe the method for creating networks with prescribed types of degree assortativity,
and in Sec.~\ref{sec:imple} we discuss aspects of the numerical implementation of the
reduced model. Results are given in Sec.~\ref{sec:res} and we conclude with a discussion
in Sec.~\ref{sec:disc}. Appendix~\ref{sec:app} contains the algorithms we use to generate
networks with prescribed assortativity.

\section{Model description and simplifications}
\label{sec:model}
We consider a network of $N$ theta neurons:
\be
   \frac{d\theta_j}{dt}=1-\cos{\theta_j}+(1+\cos{\theta_j})(\eta_j+I_j) \label{eq:dthetadt}
\ee
for $j=1,2,\dots N$ where
\be
   I_j=\frac{K}{\mk}\sum_{n=1}^NA_{jn}P_q(\theta_n) \label{eq:I}
\ee
$\eta_j$ is a constant current entering the $j$th neuron, randomly chosen from a distribution
$g(\eta)$, $K$ is strength of coupling,
$\mk$ is mean degree of the network, and the connectivity of the network is given by
the adjacency matrix $A$, where
$A_{jn}=1$ if neuron $n$ connects to neuron $j$, and zero otherwise.
The connections within the network are either all excitatory (if $K>0$) or inhibitory
(if $K<0$). Thus we do not consider the more realistic and general case of a connected
population of both excitatory and inhibitory neurons, although it would be possible
using the framework below. 

The theta neuron is the normal
form of a Type I neuron which undergoes a saddle-node on an invariant circle bifurcation (SNIC) as the
input current is increased through zero~\cite{erm96,ermkop86}. A neuron is said to fire when $\theta$
increases through $\pi$, and the function
\be
   P_q(\theta)=a_q(1-\cos{\theta})^q; \qquad q\in\{2,3,\dots\} \label{eq:pq}
\ee
 in~\eqref{eq:I} is meant
to mimic the current pulse injected from neuron $n$ to any postsynaptic neurons when
neuron $n$ fires. $a_q$ is a normalisation constant such that $\int_0^{2\pi}P_q(\theta)d\theta=2\pi$
independent of $q$.

The in-degree of neuron $j$ is defined as the number of neurons which connect to it, i.e.
\be
  k_j^{in}=\sum_{n=1}^N A_{jn}
\ee
while the out-degree of neuron $n$ is the number of neurons it connects to, i.e.
\be
  k_n^{out}=\sum_{j=1}^N A_{jn}
\ee
Since each edge connects two neurons, we can define the mean degree
\be
   \mk=\frac{1}{N}\sum_{j=1}^N k_j^{in}=\frac{1}{N}\sum_{n=1}^N k_n^{out}=\frac{1}{N}\sum_{j=1}^N\sum_{n=1}^N A_{jn}
\ee
Networks such as~\eqref{eq:dthetadt} have been studied by others~\cite{lukbar13,chahat17,lai14A},
and note that under the transformation $V=\tan{(\theta/2)}$ the theta
neuron becomes the quadratic integrate-and-fire neuron with infinite
thresholds~\cite{monpaz15,latric00}.

\subsection{An infinite ensemble}
As a first step we consider an infinite ensemble of networks with the same connectivity,
i.e.~the same $A_{jn}$, but in each member of the ensemble, the value of $\eta_j$ associated
with the $j$th neuron is randomly chosen from the distribution $g(\eta)$~\cite{barant11}. 
Thus we expect
a randomly chosen member of the ensemble to have values of $\eta$ in the ranges
\begin{align}
   \eta_1 & \in [\eta_1',\eta_1'+d\eta_1'] \nonumber \\
   \eta_2 & \in [\eta_2',\eta_2'+d\eta_2'] \nonumber \\
   & \vdots \\
   \eta_N & \in [\eta_N',\eta_N'+d\eta_N'] \nonumber
\end{align}
with probability $g(\eta_1')g(\eta_2')\dots g(\eta_N')d\eta_1'd\eta_2'\dots d\eta_N'$.
The state of this member of the ensemble is described by the probability density
\be
   f(\theta_1,\theta_2,\dots,\theta_N;\eta_1,\eta_2,\dots \eta_N;t)
\ee
which satisfies the continuity equation
\be
   \frac{\partial f}{\partial t}= -\sum_{j=1}^N\frac{\partial}{\partial\theta_j}\left\{\left[1-\cos{\theta_j}+(1+\cos{\theta_j})\left(\eta_j+I_j)\right]f\right)\right\}  \label{eq:cont}
\ee
If we define the marginal distribution for the $j$th neuron as
\be
   f_j(\theta_j,\eta_j,t)=\int f(\theta_1,\theta_2,\dots \theta_N;\eta_1,\eta_2,\dots \eta_N;t)\prod_{k\neq j}d\theta_k d\eta_k
\ee
we can write
\be
   I_j(t)
  =  \frac{K}{\langle k\rangle}\sum_{n=1}^NA_{jn} \int_{-\infty}^\infty\int_0^{2\pi} P_q(\theta_n) f_n(\theta_n,\eta_n,t)d\theta_n d\eta_n \label{eq:Ij}
\ee
where we have now evaluated $I_j$ as an average over the ensemble rather than from a single
realisation (as in~\eqref{eq:I}). This is reasonable in the limit of large networks~\cite{barant11}.

Multiplying~\eqref{eq:cont} by $\prod_{k\neq j}d\theta_k d\eta_k$ and integrating we obtain
\be
   \frac{\partial f_j}{\partial t}=  -\frac{\partial}{\partial\theta_j}\left\{\left[1-\cos{\theta_j}+(1+\cos{\theta_j})\left(\eta_j+I_j\right)\right]f_j\right\}  \label{eq:contj}
\ee
A network of theta neurons is known to be amenable to the use of the Ott/Antonsen
ansatz~\cite{ottant08,lukbar13,lai14A} so we write
\be
   f_j(\theta_j,\eta_j,t)=\frac{g(\eta_j)}{2\pi}\left[1+\sum_{k=1}^\infty \{\alpha_j(\eta_j,t)\}^k e^{ik\theta_j}+\sum_{k=1}^\infty \{\bar{\alpha}_j(\eta_j,t)\}^k e^{-ik\theta_j}\right].
\ee
The dependence on $\theta_j$ is written as a Fourier series where the $k$th coefficient
is the $k$th power of a function $\alpha_j$.
Substituting this into~\eqref{eq:contj} and~\eqref{eq:Ij}
we find that $\alpha_j$ satisfies
\be
   \frac{\partial\alpha_j}{\partial t}=-i\left[\frac{\eta_j+I_j-1}{2}+(1+\eta_j+I_j)\alpha_j+\left(\frac{\eta_j+I_j-1}{2}\right)\alpha_j^2\right] \label{eq:dalpdt}
\ee
and
\be
   I_j(t)
  =  \frac{K}{\langle k\rangle}\sum_{n=1}^NA_{jn} \int_{-\infty}^\infty H(\alpha_n(\eta_n,t);q) d\eta_n \label{eq:Ij2}
\ee
where
\be
   H(\alpha;q)=a_q\left[C_0+\sum_{n=1}^q C_n(\alpha^n+\bar{\alpha}^n)\right]
\ee
where an overbar indicates complex conjugate, and
\be
   C_n=\sum_{k=0}^q\sum_{m=0}^k\frac{\delta_{k-2m,n}q!(-1)^k}{2^k(q-k)!m!(k-m)!}
\ee
Assuming that $g$ is a Lorentzian:
\be
   g(\eta)=\frac{\Delta/\pi}{(\eta-\eta_0)^2+\Delta^2}
\ee
we can use contour integration to evaluate the integral in~\eqref{eq:Ij2}, and
evaluating~\eqref{eq:dalpdt} at the appropriate pole of $g$ we obtain
\be
  \frac{dz_j}{dt}=\frac{-i(z_j-1)^2}{2}+\frac{(z_j+1)^2}{2}\left[-\Delta+i\eta_0+iJ_j\right] \label{eq:ens}
\ee
where
\be
    J_j=\frac{K}{\langle k \rangle}\sum_{n=1}^N A_{jn}H(z_n;q) \label{eq:ensJ}
\ee
and $z_j=\langle e^{i\theta_j} \rangle$, where the expected value is taken over the ensemble.

Now~\eqref{eq:ens} is a set of $N$ coupled complex ODEs, so we have not simplified the original
network~\eqref{eq:dthetadt} in the sense of decreasing the number of equations to solve.
However, the states of interest are often {\em fixed points} of~\eqref{eq:ens}
(but not of~\eqref{eq:dthetadt}), and can
thus be found and followed as parameters are varied.
At this point the network we consider, with connectivity given by $A$, is arbitrary.
If $A$ was a circulant matrix, for example, this would represent a network of neurons
on a circle, where the strength of coupling between neurons depends only on the distance
between them~\cite{lai14A}.

\subsection{Lumping by degree}

The next step is to assume that for a large enough network, the dynamics of neurons with
the same degrees will behave similarly~\cite{chahat17}. Such an assumption has been
made a number of times in the past~\cite{kahsok17,ich04,resott14}. We thus associate
with each neuron the degree
vector ${\bm k}=(k^\inn,k^\out)$ and assume that the value of $z$ for all neurons with
a given ${\bm k}$ are similar. There are $N_{\bm{k}}=N_{k^\inn}  N_{k^\out}$ distinct degrees
where $N_{k^\inn}$ and $N_{k^\out}$ are the number of distinct in- and out-degrees, 
respectively.
We define $b_s$ to be the order parameter for neurons
with degree ${\bm k}_s$, where $s\in[1,N_{\bm{k}}]$,
and now derive equations for the evolution of the $b_s$.

Let $\bm{z}$ be the vector of ensemble states $z_j$, where $j \in [1, N]$ and the degree index of neuron $j$ be $d(j)$, such that $\bm{k}_{d(j)}$ is its degree.
We assume that for all neurons with the same degree $\bm{k}_{d(j)} = \bm{k}_s$ the ensemble state $z_j$ is similar in sufficiently large networks and thus we only care about the mean value $\mean{z_j}_{d(j) = s} = b_s$ with $s\in[1,N_{\bm{k}}]$.
We say that degree $\bm{k}_s$ occurs $h_s$ times and thus write
\begin{align}
    \bm{b} = C \bm{z},
    \label{eq:DegreeEnsembleTransformation}
\end{align}
where the $N_{\bm{k}} \times N$ matrix $C$ has $h_s$ entries in row $s$, each
 of value $1/h_s$, at positions $j$ where $d(j) = s$ and zeros elsewhere, i.e. $C_{s j} =  \delta_{s, d(j)}/h_s$ with $\delta$ being the Kronecker delta.

To find the time derivative of $\bm{b}$ we need to express $\bm{z}$ in terms of $\bm{b}$, which
we do with an $N \times N_{\bm{k}}$ matrix $B$ 
which assigns to $z_j$ the corresponding $b_s$ value, such that
\begin{align}
    \bm{z} = B \bm{b},
    \label{eq:DegreeEnsembleTransformation2}
\end{align}
with components $B_{j s} = \delta_{d(j),s}$.
Note that $CB = I_{N_{\bm{k}}}$, the $N_{\bm{k}} \times N_{\bm{k}}$ identity matrix.
Differentiating~\eqref{eq:DegreeEnsembleTransformation} with respect to time, 
inserting~\eqref{eq:ens} into this and writing $\bm{z}$ in terms of $\bm{b}$ 
using~\eqref{eq:DegreeEnsembleTransformation2} we obtain
\begin{align}
    \Dot{b_s} = \underbrace{\sum_{j=1}^N C_{sj} \left[ - i \frac{\left(\sum_{t=1}^{N_{\bm{k}}} B_{j t} b_{t}-1\right)^2}{2}  + i \frac{\left(\sum_{t=1}^{N_{\bm{k}}} B_{jt} b_{t}+1\right)^2}{2} \left(\eta_0 + i\Delta \right) \right]}_{\text{local}} \notag \\
    +
\underbrace{\sum_{j=1}^N C_{sj} \left[i \frac{\left(\sum_{t=1}^{N_{\bm{k}}} B_{j t} b_{t}+1\right)^2}{2} J_j\right]}_{\text{non-local}} . \label{eq:dbs}
\end{align}

Considering that for all $t$ there is only a single non-zero entry $B_{j t}$, equal to 1, the identity
\begin{align}
    \left(\sum_{t=1}^{N_{\bm{k}}} \underbrace{B_{j t}}_{=\delta_{d(j),t}} b_{t}\right)^n &= b_{d(j)}^n
    \label{eq:TransformationIdentity}
\end{align}
holds for any power $n$.
Further we find that
\begin{align}
    \sum_{j=1}^N \underbrace{C_{sj}}_{= 1/h_s \cdot \delta_{s, d(j)}} b_{d(j)} = b_s.
\end{align}
Thus, the local term in~\eqref{eq:dbs} is
\begin{align}
    \Dot{b_s}^\text{local} = - i \frac{\left(b_s-1\right)^2}{2}  + i \frac{\left(b_s+1\right)^2}{2} \left(\eta_0 + i\Delta \right).
\end{align}
For the non-local term we write
\begin{align}
    \Dot{b_s}^\text{non-local} &= \sum_{j=1}^N \frac{1}{h_s} \delta_{s, d(j)} i \frac{\left(b_{d(j)}+1\right)^2}{2} J_j \\
    &= i \frac{\left(b_s+1\right)^2}{2} \underbrace{\sum_{j=1}^N \frac{1}{h_s} \delta_{s, d(j)} J_j}_{= \sum_{j=1}^N C_{s j} J_j = \tilde{J}_s }, \nonumber
\end{align}
where $\tilde{J}_s$ describes the synaptic current of the ensemble equations averaged over nodes sharing the same degree $\bm{k}_s$.
The identity~\eqref{eq:TransformationIdentity} also applies to~\eqref{eq:ensJ}, so  that
\begin{align}
    H(z_n; q) &= H\left(\sum_{t=1}^{N_{\bm{k}}}B_{nt}b_t; q\right)=\sum_{t=1}^{N_{\bm{k}}}B_{nt} H(b_t; q) 
\end{align}
and the current can be written as
\begin{align}
    \tilde{J}_s &=   \sum_{j=1}^N C_{s j} \frac{K}{\mean{k}} \sum_{n=1}^{N} A_{j n} \sum_{t=1}^{N_{\bm{k}}} B_{n t} H(b_t; q) \nonumber \\
    &= \frac{K}{\mean{k}} \sum_{t=1}^{N_{\bm{k}}} \underbrace{\sum_{j=1}^N \sum_{n=1}^{N} C_{s j}  A_{j n} B_{n t}}_{E_{st}} H(b_t; q)
\end{align}
The effective connectivity between neurons with different degrees is therefore 
expressed in the matrix $E=CAB$ and we end up with equations governing the $b_s$:
\be
  \frac{db_s}{dt}=\frac{-i(b_s-1)^2}{2}+\frac{(b_s+1)^2}{2}\left[-\Delta+i\eta_0+i\tilde{J}_s\right] \label{eq:dbdt}
\ee
where
\be
    \tilde{J}_s=\frac{K}{\mean{k}}\sum_{t=1}^{N_{\bm{k}}} E_{st} H(b_t;q)  \label{eq:Jb}
\ee
These equations are of the same form as~\eqref{eq:ens}-\eqref{eq:ensJ} except that $A$ has been replaced by $E$. Note that the connectivity matrix $A$ is completely general; we have only
assumed that neurons with the same degrees behave in the same way.
We are not aware of a derivation of this form being previously presented.

\section{Network assembly}
\label{sec:assort}
We are interested in the effects of degree assortativity on the dynamics of the network of neurons.
We will choose a default network with no assortativity and then introduce one of the four types of assortativity and investigate the changes in the network's dynamics. 
Our default network is of size $N=5000$ neurons where in- and out-degrees $k$ for each neuron are independently drawn from the interval $[750, 2000]$ with probability $P(k)\sim k^{-3}$
(i.e.~a power law, as found in~\cite{eguchi05} and used in~\cite{chahat17}).
%We create such a network using the configuration model~\cite{new03}. 
We create networks using the configuration model~\cite{new03}, then modify them using
algorithms which introduce assortativity and then remove multiple connections between nodes (or multi-edges) (described in Appendix~\ref{sec:app}). Further, to observe any influence of multi-edges on the dynamics investigated here, we also developed a novel network assembly technique permitting introduction of very high densities of multi-edges also described in the Appendix; we refer to this novel technique as the ``permutation'' method.  We choose as our default parameters 
$\eta_0=-2,\Delta=0.1,K=3$, for which
%We choose as our default parameters $\eta_0=-2,\Delta=0.1,K=3$, 
 a default network approaches a stable fixed point.
The sharpness of the synaptic pulse function is set to $q=2$ for all simulations.

\begin{figure}
    \includegraphics[width=.8\linewidth]{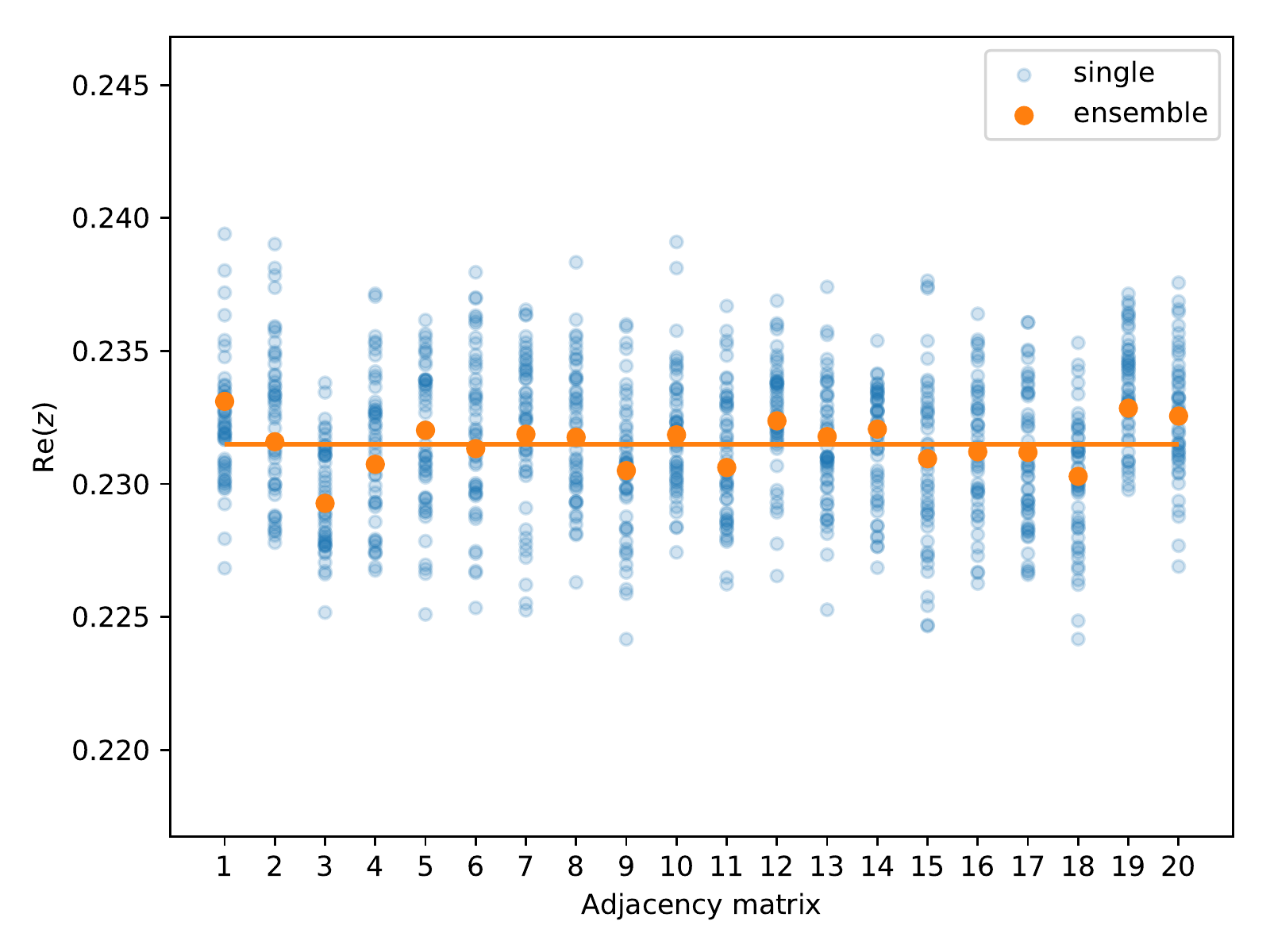}
    \caption{Orange circles: steady state of~\eqref{eq:ens}-\eqref{eq:ensJ} for 20
different default networks.
Blue circles: results from 50 different realisations of the $\eta_i$ 
for~\eqref{eq:dthetadt}-\eqref{eq:I}, for each network.  
Parameters: $\eta_0=-2,\Delta=0.1,K=3$. The orange line marks the ensemble mean value.}
    \label{fig:ensemble}
\end{figure}

We first check the validity of averaging over an infinite ensemble. We assemble 20 different
default networks and for each, run~\eqref{eq:ens}-\eqref{eq:ensJ} to a steady
state and calculate the order parameter $z$, the mean of $B\bm{b}$. 
The real part of $z$ is plotted in orange in Fig.~\ref{fig:ensemble}.
For each of these networks we then generated 50 realisations of the $\eta_i$'s and
ran~\eqref{eq:dthetadt}-\eqref{eq:I} for long enough that transients had decayed, and then
measured the corresponding order parameter for the network of individual neurons
\be
   R = \frac{1}{N}\sum_{j=1}^N e^{i\theta_j}
\ee
and plotted its real part in blue in Fig.~\ref{fig:ensemble}. Note that the orange circles
always lie well within the range of values shown in blue.
The fact that deviations within the 50 realisations are small relative to the value
obtained by averaging over an infinite ensemble 
provide evidence for the validity of this approach, at least for these parameter values.

\begin{figure}
    \includegraphics[width=.8\linewidth]{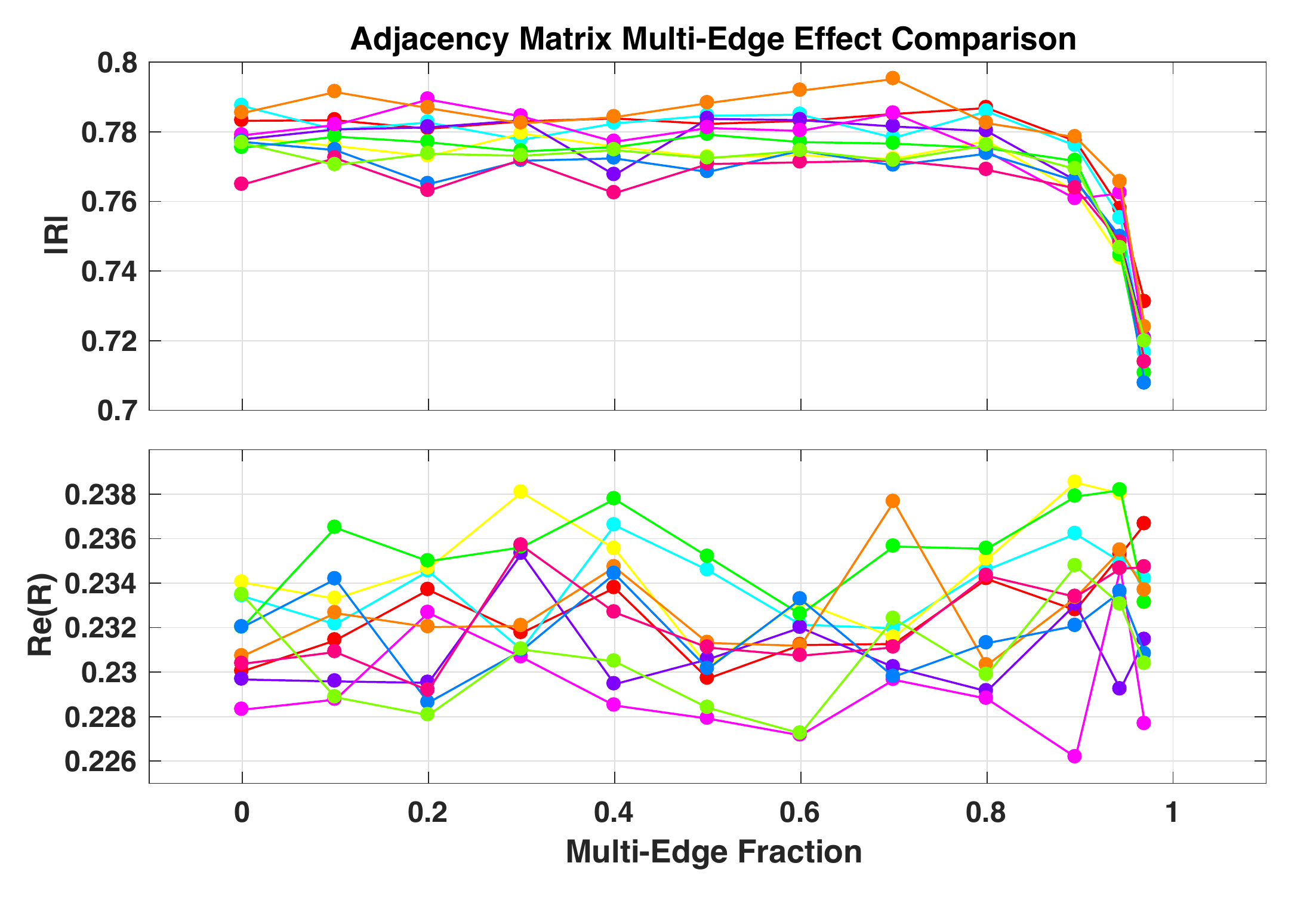}
    \caption{Comparison of steady-state values of the order parameter (top: magnitude of $R$; 
bottom: $\mbox{Re}(R)$) over a suite of adjacency matrices with varied densities of multi-edge connections ranging from none to 97\%. Higher densities of multi-edges were obtained, but assortativities exceeded the target neutral values of $\pm 0.005$. Values shown are from 
simulations of~\eqref{eq:dthetadt}-\eqref{eq:I} after initial transients decay (i.e.,~time t $\geq$ 40). Each of these 10 curves correspond to a unique realisation of default $\eta$s 
from the distribution $g(\eta)$. Parameters: $N=5000,\eta_0=-2,\Delta=0.1,K=3,q=2$.}
    \label{fig:multi-edges}
\end{figure}

We also investigate the influence of multi-connections (i.e.~more than one connection)
between neurons on the network dynamics.
The configuration model creates a network in which the neuron degrees are exactly those
specified by the choice from the appropriate distribution, but typically results in both 
self-connections and multiple connections between neurons. We have an algorithm for
systematically removing such connections while preserving the degrees, and found
that removing such edges has no significant effect (results not shown).
We also have an algorithm (see ``Permutation Method'' in Appendix~\ref{sec:app}) 
for {\em increasing} the number of multi-edges
from the number found using the default configuration model. This novel network assembly method meets specified neuron degrees and also produces specified densities of multi-connections ranging from none to 97\%; see
Fig.~\ref{fig:multi-edges} for the results of such calculations. 
We see that only when the fraction of multi-edges approaches $90\%$ do we see a significant effect.
%and we show the
%effect of removing such edges in Fig.~\ref{fig:multi-edges}. We create 20 default
%adjacency matrices and run~\eqref{eq:ens}-\eqref{eq:ensJ} to a steady state, keeping
%$100\%$ of all multi-edges. We then remove some fraction of initial multi-edges and repeat
%the process, continuing until no multi-edges remain. The real part of the order parameter
%for all cases is shown in Fig.~\ref{fig:multi-edges}, and we see that (for these parameters)
%variations between the default matrices are greater than those caused by removing all
%multi-edges. 
However, in our simulations we use simple graphs without multi-edges.

%SHAWN, PLEASE REDO FIG 3. PLOT REAL(R) ON VERTICAL AXIS, AS R IS THE NETWORK EQUIVALENT
%OF THE ORDER PARAMETER (REMOVE TOP PANEL).

%\begin{figure}
%    \includegraphics[width=.8\linewidth]{multi-edges.pdf}
%    \caption{20 different default adjacency matrices (indicated by different colors) are created,
%and then multi-edges are systematically removed. Top: real part of the order parameter at steady state. Bottom: difference of real part of $z$ from that obtained before multi-edges are removed. There is no trend observable while removing multi-edges. Parameters: $\eta_0=-2,\Delta=0.1,K=3$.}
%    \label{fig:multi-edges}
%\end{figure}

\subsection{Assortativity}

For a given matrix $A$ we can measure its assortativity by calculating the four Pearson correlation coefficients $r(\alpha, \beta)$ with $\alpha, \beta \in [\inn, \out]$ which read
\begin{align}
    r(\alpha, \beta) = \frac{\sum_{e=1}^{N_e} ({}^{s}k^{\alpha}_e - \left<{}^{s}k^\alpha \right> )({}^{r}k^{\beta}_e - \left<{}^{r}k^\beta \right> )}{\sqrt{\sum_{e=1}^{N_e} ({}^{s}k^{\alpha}_e - \left<{}^{s}k^\alpha \right> )^2} \sqrt{\sum_{e=1}^{N_e} ({}^{r}k^{\beta}_e - \left<{}^{r}k^\beta \right> )^2} }
    \label{eq:AssortativityDefinition}
\end{align}
where
\begin{align}
    \left<{}^{s}k^\alpha \right> = \frac{1}{N_e} \sum_{e=1}^{N_e} {}^{s}k^{\alpha}_e && \text{and} && \left<{}^{r}k^\beta \right> = \frac{1}{N_e} \sum_{e=1}^{N_e} {}^{r}k^{\beta}_e,
    \label{eq:AssortativityMeanDefinition}
\end{align}
$N_e$ being the number of connections and the leading superscript $s$ or $r$ refers to the sending or receiving neuron of the respective edge.
For example the sending node's in-degree of the second edge would be ${}^{s}k^{in}_2$.
Note that there are four mean values to compute.

We introduce assortativity by randomly choosing two edges and swapping
postsynaptic neurons when doing so would increase the target assortativity
coefficient~\cite{schkih15}.
An edge $(i, j)$ is directed from neuron $j$ to neuron $i$. In order to know whether the pair $(i,j)$ and $(h,l)$ should be rewired or left untouched, we compare their contribution to the covariance in the numerator of~\eqref{eq:AssortativityDefinition}:
\begin{align}
    c_{\parallel} &= c((i,j), (h,l)) \notag\\
    &= \left(k^{\alpha}_j - \left<{}^{s}k^\alpha \right> \right)\left(k^{\beta}_i - \left<{}^{r}k^\beta \right> \right) +
    \left(k^{\alpha}_l - \left<{}^{s}k^\alpha \right> \right)\left(k^{\beta}_h - \left<{}^{r}k^\beta \right> \right);\\
    c_{\crossed} &= c((i,l), (h,j))\notag\\
    &= \left(k^{\alpha}_l - \left<{}^{s}k^\alpha \right> \right)\left(k^{\beta}_i - \left<{}^{r}k^\beta \right> \right) +
    \left(k^{\alpha}_j - \left<{}^{s}k^\alpha \right> \right)\left(k^{\beta}_h - \left<{}^{r}k^\beta \right> \right).
\end{align}
If $c_{\crossed}>c_{\parallel}$ we replace the edges $(i,j)$ and $(h,l)$ by $(i,l)$ and $(h,j)$,
respectively, 
otherwise we do not, and continue by randomly choosing another pair of edges. Algorithm \ref{alg:assortative_mixing} (see Appendix~\ref{sec:app}) demonstrates a scheme for reaching a certain target assortativity coefficient.

We investigate the effects of different types of assortativity (see 
Fig~\ref{fig_assort_type_schematic1}) in isolation.
We thus need a family of networks parametrised by the relevant assortativity coefficient.
Algorithm~\ref{alg:assortative_mixing} is used to create a network with a specific value
of one of the assortativity coefficients, but especially for high values of assortativity
it may be that in doing so a small amount of assortativity of a type other than the intended one is introduced. Accordingly, it may be necessary to examine all types 
of assortativity and apply the mixing scheme to reduce other types back to zero, and then
(if necessary) push the relevant value of assorativity back to its target value.
We do multiple iterations of these mixing rounds until all assortativities are at 
their target values (which may be 0) within a range of $\pm0.005$.
We use Algorithm~\ref{alg:assortative_mixing} with a range of target assortativities $r$, and for
each value, store the connectivity matrix $A$ and thus form the parametrised family $E(r)$.
We do this for the four types of assortativity.

%
%\textcolor{blue}{Rewrite above paragraph:}
%Investigating the influence of each assortativity coefficient calls for whole families of networks spanning ranges of values for each individual assortativity type (see Fig \ref{fig_assort_type_schematic1}). By application of our Algorithm \ref{alg:assortative_mixing}, we can store $A$ at various points of the algorithm's progress while increasing or decreasing the respective $r$ of interest; we thus form parameterised families of $E(r)$. Note, inspection of each assortativity type in isolation requires only the targeted $r$ pushed outside our `default' parameter regime, or, $r \notin [-0.005,0.005]$. This entails, however, iteratively and selectively manipulating the various assortativities over multiple applications simply due to some $r$ values requiring decreases back to zero while we elevate the targeted $r$ of interest. Despite this complication, and limitations of precision achieved with targeted $r$ values (e.g., swapping edges can only achieve so much precise effect on $r$), we nevertheless maintain accuracies for desired $r$ values within a range of $\pm0.005$.

We have chosen to use the configuration model to create networks with given degree sequences and then introduced assortativity by swapping edges. Although we developed our novel ``permutation'' method as well (see Appendix~\ref{sec:app}), that method was designed for assembling adjacency matrices with desired multi-edge densities and was applied only for that aspect. By contrast, another common adjacency network assembly technique, that of Chung and Lu~\cite{chulu02} together with an analytical expression for assortativity
(as in~\cite{chahat17}), proved inadequate. We found that the latter approach significantly alters the degree distribution
for large assortativity, whereas the configuration model combined with our mixing algorithm does not change degrees at all. For our default network this approach allows us to introduce assortativity of any one kind up to $r=\pm 0.5$.

%We have chosen to use the configuration model to create networks with given degree sequences
%and then introduced assortativity by swapping edges. Another approach would be that of
%Chung and Lu~\cite{chulu02} together with an analytical expression for assortativity
%(as in~\cite{chahat17}). However,
%we found that the latter approach significantly alters the degree distribution
%for large assortativity, whereas the configuration model combined with our mixing algorithm does not change degrees at all. For our default network this approach allows us to introduce assortativity of any one kind up to $r=\pm 0.5$.

\section{Implementation}
\label{sec:imple}
For networks of the size we investigate 
it is impractical to consider each distinct in- and out-degree
(because $E$ will be very large and sparse). Due to the smoothness of the degree dependency of $b(\bm{k})$ we coarse-grain in degree space by introducing ``degree clusters'' --- lumping
all nodes with a range of degrees into a group with dynamics
described by a single variable.
Let there be $N_{c^{\inn}}$ clusters in in-degree and $N_{c^{\out}}$ clusters in out-degree, with a total of
$N_{\bm{c}}=N_{c^{\inn}} \cdot N_{c^{\out}}$ degree clusters.
The matrix $C$ then is an $N_{\bm{c}} \times N$ matrix and constructed as previously, except that $d(j)$ is not the degree index of neuron $j$, but the cluster index and $s$ is the cluster index running from 1 to $N_{\bm{c}}$. Similarly for the matrix $B$.
There are multiple options for how to combine degrees into a cluster.
The cluster index of a neuron can be computed linearly, corresponding to clusters
of equal size in degree space. However, with this approach,
depending on the degree distribution, some of the clusters may be empty or hardly filled, resulting in poor statistics.
To overcome this issue, the cumulative sum of in- and out-degree distribution can be used to map degrees to cluster indices.
Thus, clusters are more evenly filled and at the same time regions of degree space
 with high degree probability are more finely sampled.
The dynamical equations~\eqref{eq:dbdt}-\eqref{eq:Jb} are equally valid for
describing degree cluster dynamics with $s, t \in [1, N_{\bm{c}}]$ and $E=CAB$, where $C$ and $B$ are cluster versions of their previous definitions.

\begin{figure}
    \includegraphics[width=.8\linewidth]{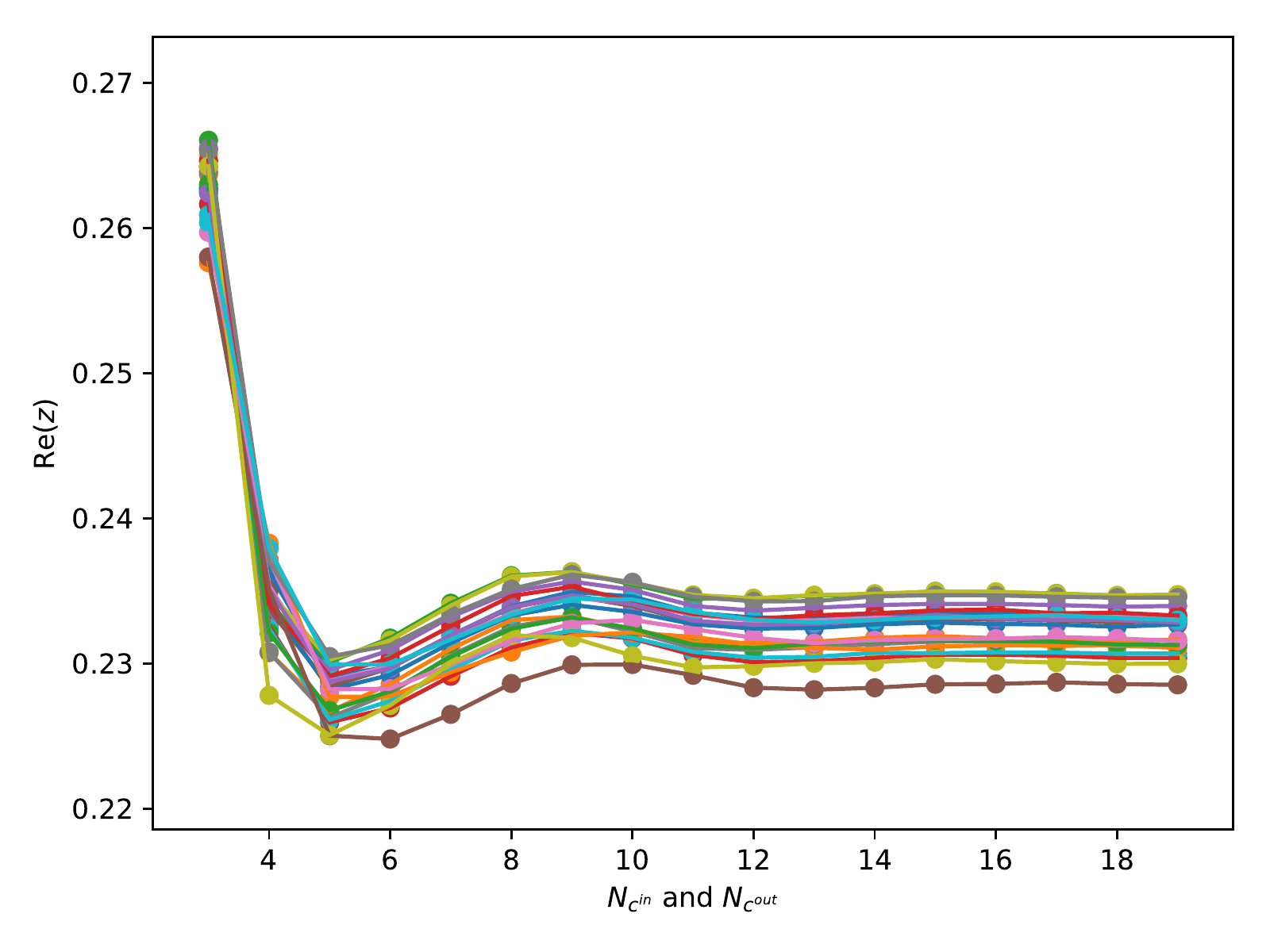}
    \caption{Real part of $z$ at steady state for 20 different default adjacency matrices (indicated by different colors), as the number of clusters in degree space is varied.  Parameters: $\eta_0=-2,\Delta=0.1,K=3$.}
    \label{fig:degree-cluster}
\end{figure}

To check the effect of varying the number of clusters we generate 20 default matrices
and then generate the corresponding matrix $E$ with varying numbers of clusters
($N_{c^{\inn}}$ and $N_{c^{\out}}$ are equal), then 
run~\eqref{eq:dbdt}-\eqref{eq:Jb} to a steady state and plot the real part of $z$ in
Fig.~\ref{fig:degree-cluster}. We see that 
the order parameter is well approximated using as little as about $N_{c^{\inn}} = N_{c^{\out}} = 10$ degree clusters. Beyond that,
fluctuations between different network realisations exceed the error introduced by clustering.
In our simulations we stick to the choice of 10 degree clusters per in- and out-degree space.

\begin{figure}
    \includegraphics[width=.8\linewidth]{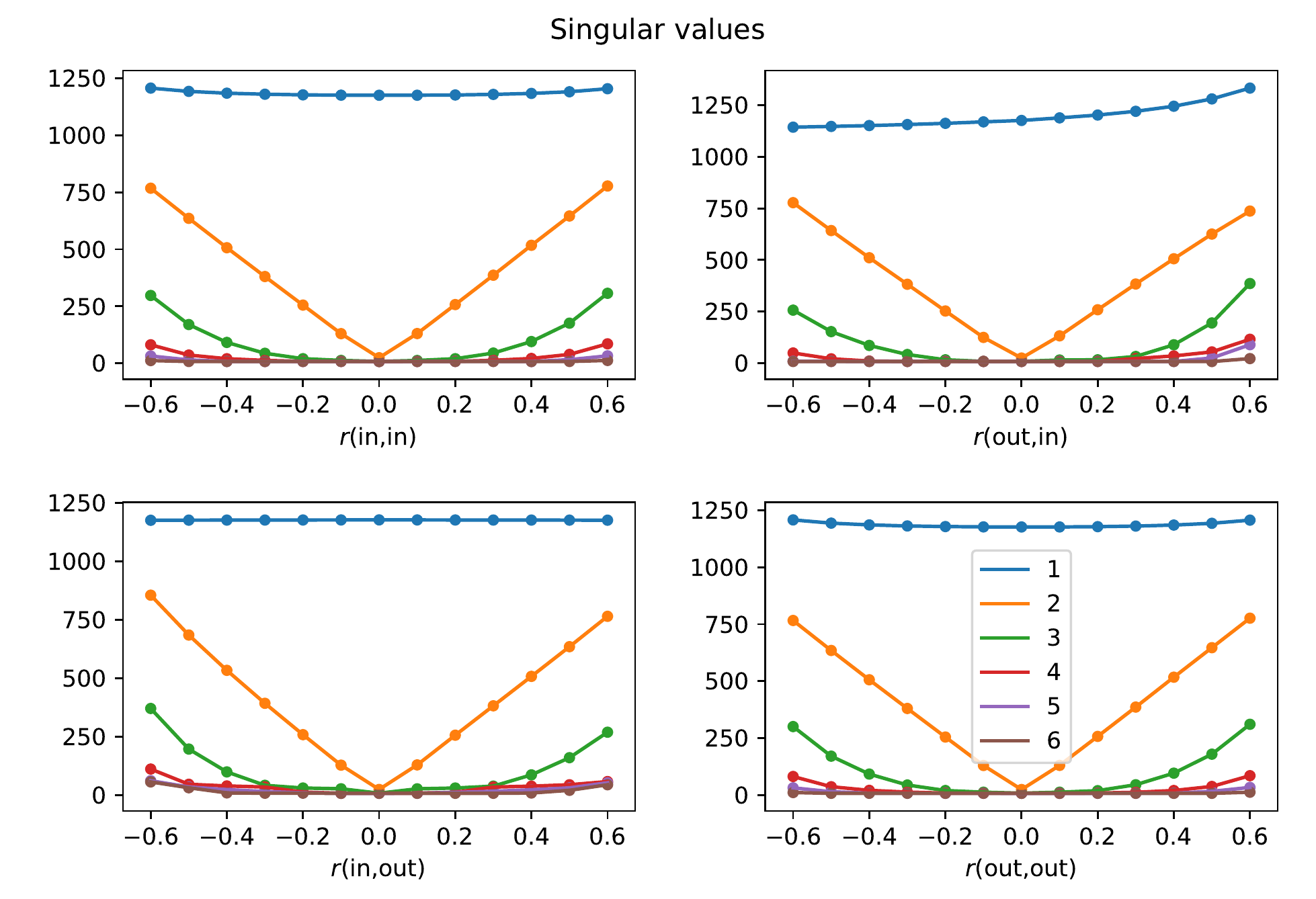}
    \caption{Six largest singular values of the SVD decomposition of $E$ as a function
of assortativity coefficent, for 4 types of assortativity.}
%decay very quickly, thus a low-rank approximation is possible. Illustrated are the six largest each in an individual color for the four assortativity types. For our simulations rank three shall be sufficient.}
\label{fig:singular-values}
\end{figure}

Having performed this clustering, we find that it is possible to represent $E$ using a low-rank approximation,
calculated using singular value decomposition. Thus for a fixed $r$ we have
\be
   E = U S V^T
\ee
where $S$ is a diagonal matrix with decreasing entries, called singular values, and
 $U$ and $V$ are unitary matrices.
In Fig.~\ref{fig:singular-values} we plot the largest 6 singular values of $E$ as function
the assortativity coefficient, for the 4 types of assortativity. Even for large $|r|$
the singular values decay very quickly, thus a low-rank approximation is possible. We choose
a rank-3 approximation, so approximate $E$ by
\be
   E(r)\approx \left[\begin{array}{ccccc} u_1(r) & \vline & u_2(r) & \vline & u_3(r)  \end{array}\right]
\begin{bmatrix} s_1(r) & 0 & 0 \\ 0 & s_2(r) & 0 \\ 0 & 0 & s_3(r) \end{bmatrix}
\left[\begin{array}{c} v_1^T(r) \\ \vspace*{-2mm} \\ \hline \\ v_2^T(r)  \\ \vspace*{-2mm} \\ \hline \\ v_3^T(r) \end{array}\right]
\ee
where $u_i$ is the $i$th column of $U$, similarly for $v_i$ and $V$, and $s_i$ is the $i$th
singular value. We have such a decomposition
at discrete values of $r$ and use cubic interpolation to evaluate $E(r)$ for any $r$.
This decomposition means that
the multiplication in \eqref{eq:Jb} can be evaluated quickly using 3 columns of $U$ and $V$
rather than the full $N_{\bm{c}} \times N_{\bm{c}}$ matrix~$E$.

We note that the components for the approximation of $E(r)$ are calculated once and then
stored, making it very easy to systematically investigate the effects of varying
any of the parameters $\eta_0,\Delta,K$ and $q$ (governing the sharpness of the
pulse function~\eqref{eq:pq}).

%We want to use numerical continuation on solutions of\ ~\eqref{eq:dbdt}-\eqref{eq:Jb} as $r$ is varied, and thus we need the approximation of $E$ to be a smooth function of $r$.
%We do this by generating adjacency matrices for a series of assortativity coefficients and compute respective $E$ matrices.
%Subsequently, we make use of the low-rank approximation and further we fit cubic polynomials through each function in $U$ and $V$ and store those coefficients and singular values $S$.
%In total that is a set of 27 variables for each assortativity coefficient and by cubic interpolation we gain such a set for an arbitrary value.

\section{Results}
\label{sec:res}

%We can not easily test whether our degree mean field theory produces exact results, since $E$ is a matrix of size $N_{\bm{k}} \times N_{\bm{k}}$, which means for the default setup $2.4\cdot10^{12}$ entries and easily several terabyte of memory.
%Therefore, we immediately introduce degree clustering and see in \autoref{fig:degree-cluster} that in fact not many clusters are necessary to reproduce the order parameter of a full network of neurons.
%
%
%For the decision on how many values to consider for the low-rank approximation, we apply SVD to $E$ and look at the largest singular values for all kinds of assortativity shown in \autoref{fig:singular-values}.
%In our simulations we use three.

\subsection{Excitatory coupling}
We take $K = 3$ to model a network with only excitatory connections.
To study the dynamical effect of assortativity we generate positive and negative ($r=\pm0.2$) assortative networks of the four possible kinds and follow fixed points 
of~\eqref{eq:dbdt}-\eqref{eq:Jb} as
a function of $\eta_0$, and compare results with a neutral ($r=0$) network.
We use pseudo-arc-length continuation~\cite{lai14B,gov00}.

To calculate the mean frequency over the network we evaluate $\bm{z}=B\bm{b}$
and then use the result that if the order parameter
at a node is $z$, then the frequency of neurons at that node is~\cite{lai15,monpaz15}
\be
   \frac{1}{\pi}\mbox{Re}\left(\frac{1-\bar{z}}{1+\bar{z}}\right)
\ee
Averaging these gives the mean frequency.

Results  are shown in \autoref{fig:excitatory_coupling}, where we see quite similar behaviour in each case: apart from a bistable region containing two stable and one unstable fixed point, there is only a single stable fixed point present.
Further, the two assortativity types (out,in) and (out,out) apparently do not affect the dynamics, whereas the saddle-node bifurcations marking the edges of the bistable region
move slightly for (in,out) and significantly for (in,in) assortativity.
Following the saddle-node bifurcations for the latter two cases we find the results shown in \autoref{fig:excitatory_coupling_full}. We have performed similar calculations for
different networks with the same values of assortativity and found similar results. 

\begin{figure}
    \includegraphics[width=.9\linewidth]{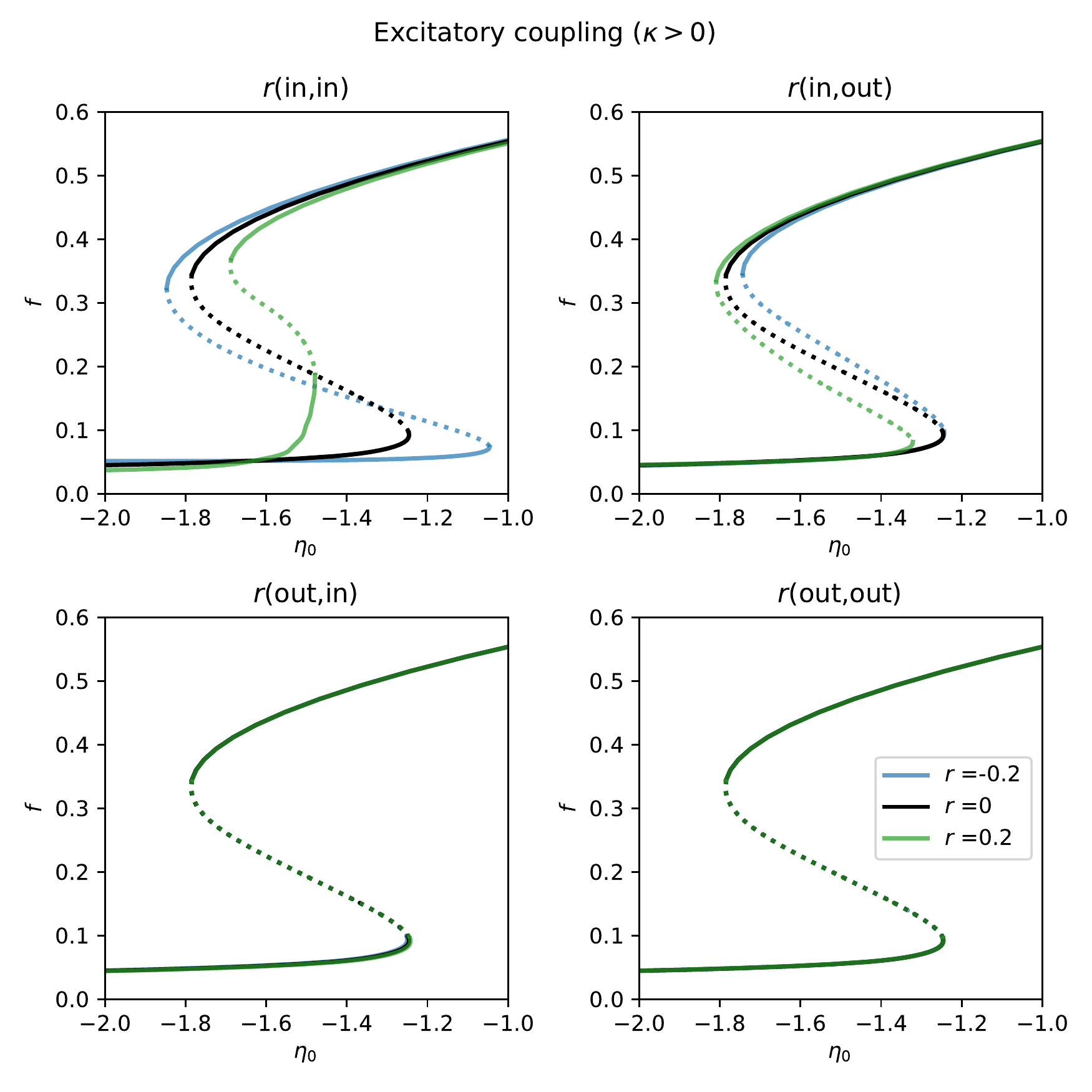}
    \caption{Average firing rate at fixed points of~\eqref{eq:dbdt}-\eqref{eq:Jb}
as a function of $\eta_0$, for the 4 types of assortativity. 
For each type of assortativity curves are plotted for $r=0$ (black), $r=-0.2$ (blue) and 
$r=0.2$ (green). Solid lines indicate stable and dashed lines unstable fixed points. Parameters: $K=3,\Delta=0.1$.}
\label{fig:excitatory_coupling}
\end{figure}

\begin{figure}
    \includegraphics[width=.9\linewidth]{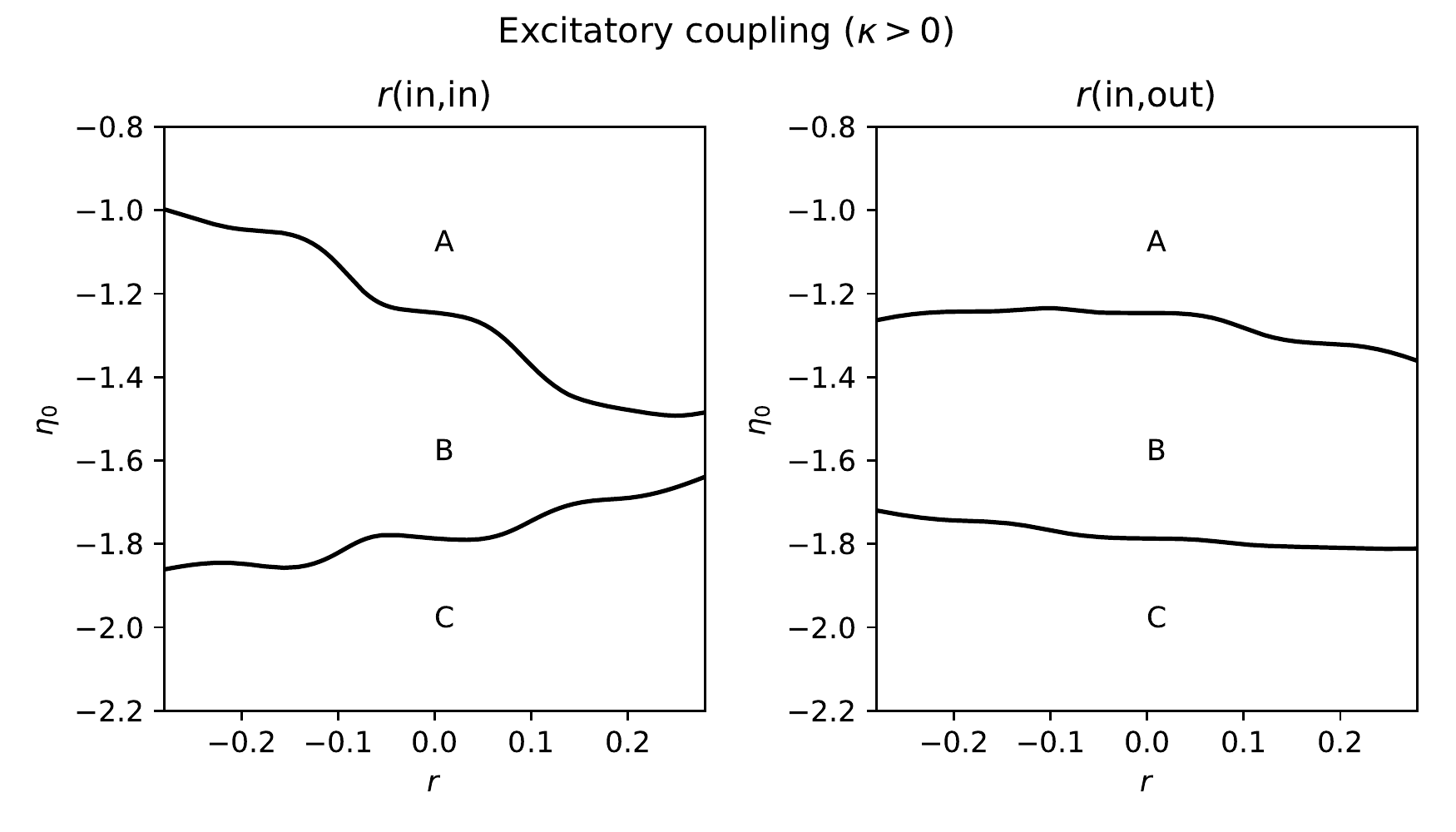}
    \caption{Continuation of the saddle-node bifurcations seen in the upper two panels of
Fig.~\ref{fig:excitatory_coupling} as $r$ is varied. Curves in \autoref{fig:excitatory_coupling} correspond to vertical slices at $r=0,\pm 0.2$. The network is bistable in region $B$ and
has a single stable fixed point in regions A and C.}
\label{fig:excitatory_coupling_full}
\end{figure}

\subsection{Inhibitory coupling}
We choose $K = -3$ to model a network with only inhibitory coupling.
Again, we numerically continue fixed points for zero, positive and negative assortativity
($r=0,\pm0.2$) as $\eta_0$ is varied and obtain the curves shown in \autoref{fig:inhibitory_coupling}.
Consider the lower left plot.
For large $\eta_0$ the system has a single stable fixed point which undergoes a supercritical
Hopf bifurcation as $\eta_0$ is decreased, creating a stable periodic orbit.
This periodic orbit is destroyed in a saddle-node bifurcation on an invariant circle (SNIC)
bifurcation at lower $\eta_0$, forcing the oscillations to stop.
Decreasing $\eta_0$ further, two unstable fixed points are destroyed in a saddle-node
bifurcation. In contrast with the case of excitatory coupling, oscillations in the
average firing rate are seen. These can be thought of as partial synchrony, since some
fraction of neurons in the network have the same period and fire at similar times to cause
this behaviour. The period of this macroscopic oscillation tends to infinity as the SNIC
bifurcation is approached, as shown in the inset of the lower left panel in  
Fig.~\ref{fig:inhibitory_coupling}.

As in the excitatory case, we see that assortativities of 
type (out,in) and (out,out) have no influence on the dynamics in this scenario.
However, type (in,out) does have a small effect, slightly moving bifurcation points
(top right panel in Fig.~\ref{fig:inhibitory_coupling}). Type (in,in) has the strongest
effect, resulting in a qualitative change in the bifurcation scenario for large enough
assortativity: there is a region of bistability between either two fixed points or
a fixed point and a periodic orbit.
This is best understood by following the bifurcations in the top panels
of Fig.~\ref{fig:inhibitory_coupling}
as $r$ is varied, as shown in \autoref{fig:inhibitory_coupling_full}.
There is one fixed point in regions A, B and D, and three in region C.
For (in,out) assortativity there is a stable periodic orbit in region B and never any
bistability.

\begin{figure}
    \includegraphics[width=.9\linewidth]{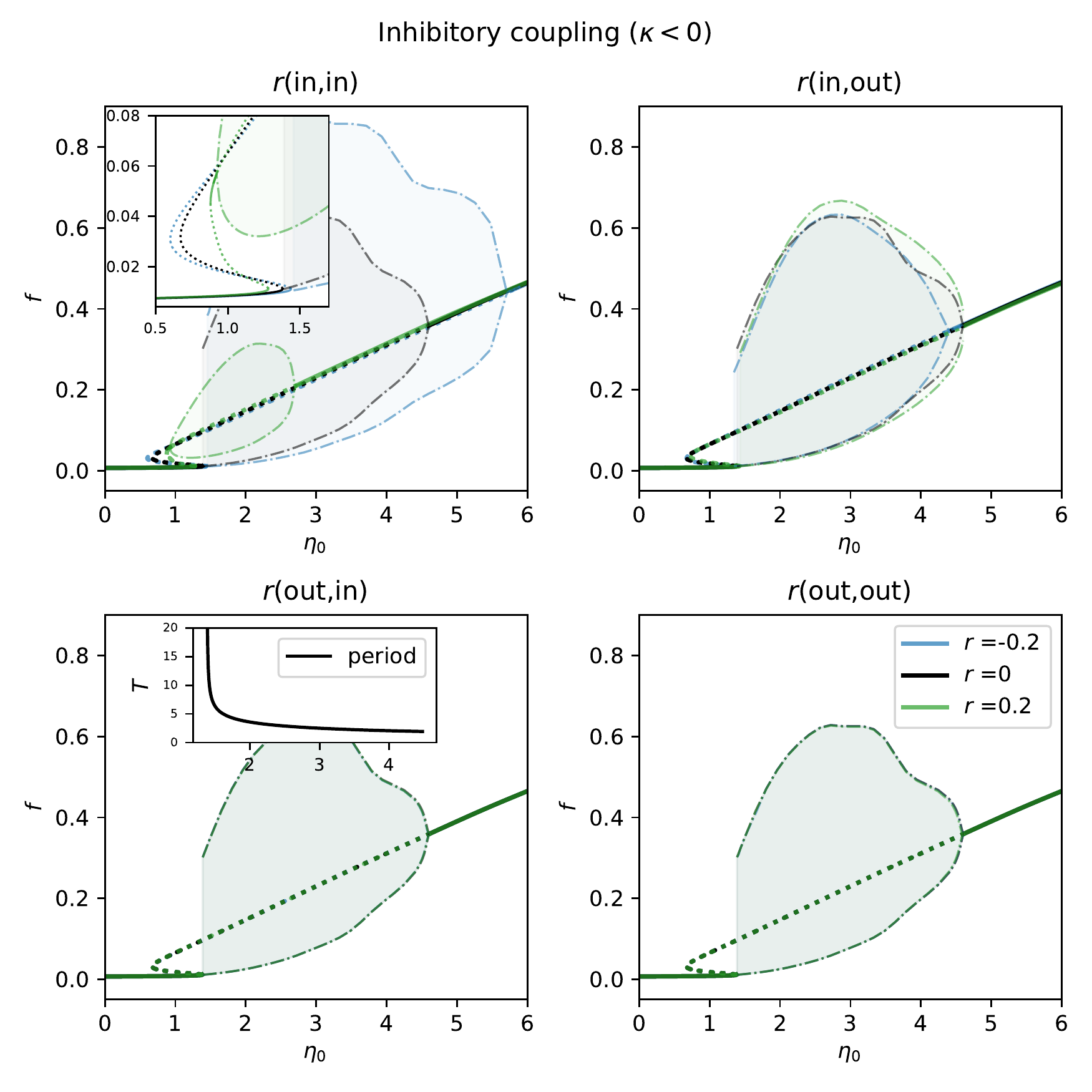}
    \caption{Average firing rate at fixed points of~\eqref{eq:dbdt}-\eqref{eq:Jb}
as a function of $\eta_0$, for the 4 types of assortativity. 
For each type of assortativity curves are plotted for $r=0$ (black), $r=-0.2$ (blue) and 
$r=0.2$ (green). In addition there are oscillations in certain regions and dash-dotted lines outline the minimal and maximal firing rate over one period of oscillation. 
The (in,in)-plot in the top left corner contains a zoom of rest of the panel, 
and the (out,in)-plot contains a subplot with the oscillation's period for $r=0$ and which is aligned with the outer $\eta_0$ axis.}
\label{fig:inhibitory_coupling}
\end{figure}

We now describe the case for (in,in) assortativity. For negative and zero $r$ the scenario
is the same as for the other three types, but as $r$ is increased there is a Takens-Bogdanov
bifurcation where regions C,D,E and F meet, 
leading to the creation of a curve of homoclinic bifurcations, which is 
destroyed at another codimension-two point where there is a homoclinic connection to a
non-hyperbolic fixed point~\cite{cholin90}. There are stable oscillations in region E,
created or destroyed in supercritical Hopf or homoclinic bifurcations. In region F there is
bistability between two fixed points.

%, where there is one fixed point in region A, B and D, which is stable in A and D, and an unstable spiral in B leading to oscillations.
\begin{figure}
    \includegraphics[width=.9\linewidth]{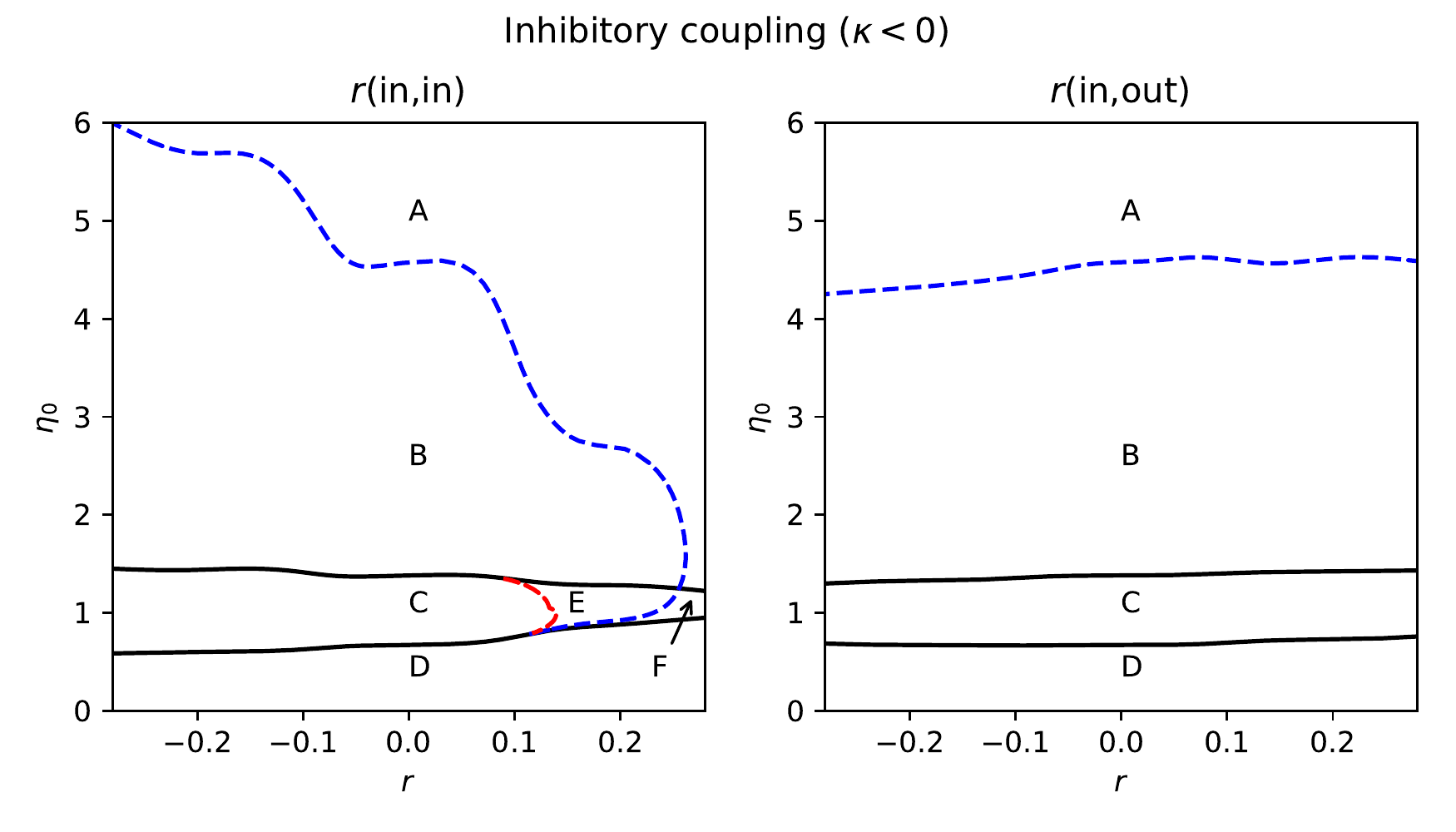}
    \caption{Continuation of bifurcations seen in upper panels of
Fig.~\ref{fig:inhibitory_coupling}.
 Solid black lines indicate saddle-node bifurcations, dashed blue is a Hopf bifurcation and dashed red a homoclinic bifurcation. Curves in \autoref{fig:inhibitory_coupling} can be understood as vertical slices through the respective plot at $r=0,\pm 0.2$. See text for explanation of labels.}
\label{fig:inhibitory_coupling_full}
\end{figure}
%Region C still contains the same unstable spiral as in B, but a SNIC has added a stable fixed point and a saddle and stopped oscillations.
%For positive (in, in) assortativity we find the saddle-node is not created on the periodic orbit anymore, allowing further oscillations in region E.
%There is a homoclinic bifurcation between C and E, where the saddle becomes part of the orbit while connecting its unstable and stable manifold.
%At a critical point homoclinic and saddle-node bifurcation unite, leading to a bifurcation of codimension two~\cite{cholin90}.
%In region F the system shows bistability, due to two stable and one unstable fixed points.
%This caused by a Hopf bifurcation occuring at the transition between E and F.
%When this Hopf bifurcation is colliding with the homoclinic and the saddle-node bifurcation, i.e. where C, D, E and F meet, the system undergoes a Takens-Bogdanov bifurcation.

%Need to check that introducing one type of assortativity does not change other
%types (pretty sure it doesn't) or other properties of the network (doesn't change
%degree distributions, or correlations between in- and out-degree of a single neuron).\text
%There are many of these \begin{verbatim}(https://en.wikipedia.org/wiki/Network_science#Network_properties) \end{verbatim}
%and it would be worth checking a few.
%\textcolor{green}{[I only checked degree distribution, degree-correlation within neurons and all 4 kinds of assortativity when mixing a matrix. Shall there be a figure or I could extent the comment at the end of 2.3 Assortativity section.]}

\section{Discussion}
\label{sec:disc}
We investigated the effects of degree assortativity on the dynamics of a network of theta neurons.
We used the Ott/Antonsen ansatz to derive evolution equations for an order parameter
associated with each neuron, and then coarse-grained by degree and then degree cluster,
obtaining a relatively small number of coupled ODEs, whose dynamics as parameters varied
could be investigated using numerical continuation. 
We found that degree assortativity involving the out-degree of the sending neuron, i.e. (out,in) and (out,out), has no effect on the networks' dynamics.
Further, (in,out) assortativity moves bifurcations slightly, but does not lead to substantial differences in dynamical behaviour.
The most significant effects were caused by creating correlation between in-degrees of the 
sending and receiving neurons.
For our excitatorially coupled example, positive (in,in) assortativity narrows the bistable region, whereas negative assortativity widens it (see Fig.~\ref{fig:excitatory_coupling_full}).
In the inhibitory case introducing negative assortativity increased the amplitude of 
network oscillations and extended their range to slightly larger $\eta_0$.
On the contrary, positive (in,in) assortativity in this network has an opposite effect and eventually stops oscillations (see Fig.~\ref{fig:inhibitory_coupling_full}).

The most similar work to ours is that of~\cite{chahat17}. These authors also considered a network
of the form~\eqref{eq:dthetadt}-\eqref{eq:I} and by assuming that the dynamics depend
on only a neuron's degree and that the $\eta_j$ are chosen from a Lorentzian,
and using the Ott/Antonsen ansatz, they derived equations similar to~\eqref{eq:dbdt}-\eqref{eq:Jb}.
The difference in formulations is that rather than a sum over entries of $E$
(in~\eqref{eq:Jb}),~\cite{chahat17} wrote the sum as
\be
   \sum_{{\bf k}'}P({\bf k}')a({\bf k}'\rightarrow {\bf k})
\ee
where $P({\bf k})$ is the degree distribution and $a({\bf k}'\rightarrow {\bf k})$ is the
assortativity function, which specifies the probability of a link from a node with degree
${\bf k}'$ to one with degree ${\bf k}$ (given that such neurons exist).
They then chose a particular functional
form for $a$ and briefly presented the results of varying one type of assortativity
(between $k_{in}'$ and $k_{out}$). In contrast, our approach is far more general
(since {\em any} connectivity matrix $A$ can be reduced to the corresponding $E$, the only
assumption being that the dynamics are determined by a neuron's degree). We also show the
results of a wider investigation into the effects of assortativity.

This alternative presentation also explains why $E$ can be well approximated with a
low-rank approximation. If the in- and out-degrees of a single neuron are independent,
$P({\bf k}')=P_i(k_{in}')P_o(k_{out}')$, and with neutral assortativity,
$a({\bf k}'\rightarrow {\bf k})=k_{out}'k_{in}/(N\mk)$. Thus
\be
   \sum_{{\bf k}'}P({\bf k}')a({\bf k}'\rightarrow {\bf k})H(b({\bf k}');q)
=\frac{k_{in}}{N\mk}\sum_{k_{out}'}\sum_{k_{in}'}k_{out}'P_i(k_{in}')P_o(k_{out}')H(b(k_{out}',k_{in}');q)
\ee
This term contributes to the input current to a neuron with degree ${\bf k}=(k_{in},k_{out})$,
but is independent of $k_{out}$. Thus the state of a neuron depend only on its in-degree, so
\be
   \sum_{{\bf k}'}P({\bf k}')a({\bf k}'\rightarrow {\bf k})H(b({\bf k}');q)=\frac{k_{in}}{N}\sum_{k_{in}'}P_i(k_{in}')H(b(k_{in}');q)
\ee
Comparing with~\eqref{eq:Jb} we see that $E={\bf c}^T {\bf d}$ where
${\bf c}=(k_{in}^1,k_{in}^2\dots k_{in}^{N_{k^{in}}})/N$ and
${\bf d}=(P_i(k_{in}^{1}),P_i(k_{in}^{2})\dots,P_i(k_{in}^{N_{k^{in}}}))$, i.e.~$E$ is a
rank-one matrix. Varying assortativity within the network is then a perturbation away
from this, with the effects appearing in the second (and third) singular values in the
SVD decomposition of $E$. 

A limitation of our study is that we
 considered only networks of fixed size with the same distributions of in- and out-degrees,
and a specific distribution of these degrees. However, our approach does not rely on this
and could easily be adapted to consider other types of networks, although we expect it
to become less valid as both the average degree and number of neurons in the network
decrease.
We have also only considered theta neurons, but since a theta neuron is the normal form
of a Type I neuron, we expect similar networks of other Type I neurons to behave
similarly to the networks considered here.  
The approach presented here could also be used to efficiently investigate the effects of correlated heterogeneity, where either the mean or width of the distribution of the $\eta_j$ is correlated with a neuron's in- or out-degree~\cite{skasun13,sonsag13,cougol13}.
We could also consider assortativity based on a neuron's intrinsic drive ($\eta_j$)~\cite{skares15} rather than its degrees, or correlations between an individual neuron's in- and out-degree~\cite{vashou12,lamsmi10,marhou16,vegrox19,nykfri17}. We are currently investigating such ideas.

{\bf Acknowledgements:} This work is partially supported by the Marsden Fund Council
from Government funding, managed by Royal Society Te Ap\={a}rangi. We thank the referees
for their helpful comments.

%\begin{figure}
%\includegraphics[scale=0.85]{degs}
%\caption{``Target'' degrees (top) and actual degrees in generated network (bottom).
%Left: in-degrees; right: out-degrees.}
%\label{fig:degs}
%\end{figure}

%\pagebreak
\appendix

\section{Algorithms}
\label{sec:app}
%\subsection{Algorithms for Adjacency Matrices}
%\textcolor{blue}{I just shifted the assortative modificaton algorithm here, added in algorithm for multi-edge removal of configuration model assemblies, and brief description of the `permutation' method.}
We present here the algorithms developed and utlised for adjacency matrix assembly and modification. These include modifications to resulting adjacency matrices produced by the familiar configuration method, and our novel matrix assembly technique we christen the ``permutation'' method.

\subsection{Assortativity}
Algorithm~\ref{alg:assortative_mixing} is used to create a network with a specified
degree distribution and values of the four different assortativity coefficients.

\begin{algorithm}
    \caption{Assortative mixing. \newline
    \textit{Randomly pair up all $N_e$ edges of the network with adjacency matrix $A$ and reconnect them at once where preferable with respect to target assortativity $r_\text{target}$. Repeat the process until the assortativity coefficient lies within the tolerance.
    Once overshooting the target coefficient, interpolate the length of a shortened list of edge pairs and reconnect those.
    } }
    \SetAlgoLined
    \label{alg:assortative_mixing}

    \tcc{compute difference in assortativity}
    $\Delta r = r_\text{target} - r(A)$\;
    \While{\upshape $|\Delta r| > tolerance$}{
        pair up all edges $[(i,j),(k,l)]$\;
        \tcc{compute whether each pair should be reconnected}
        $s_{\Delta r} = [$true: if reconnection will minimise $\Delta r$; false: otherwise$]$\label{algline:selection_criteria_r}\;
        \tcc{trial reconnection}
        $A^\ast = \operatorname{copy}(A)$\;
        reconnect edges in $A^\ast$ according to $s_{\Delta r}$\;
        $\Delta r^\ast = r_\text{target} - r(A^\ast)$\;
        \If{\upshape $\operatorname{sign}( \Delta r^\ast ) \neq \operatorname{sign}( \Delta r )$}{
            \tcc{r(A$^\ast$) is already beyond the target:}
            \tcc{limit number of edges for reconnection process}
            interpolation data $\Gamma$: $(0, r(A)), (N_e/2, r(A^\ast))$\;
            \While{$|\Delta r^\ast| > tolerance$}{
                interpolate $(L, r_\text{target})$ using $\Gamma$\;
                $s_\text{limit} = [$true: list index $< L$; false: list index $> L]$\;
                \tcc{trial selection and reconnection}
                $s^\ast = s_{\Delta r} \land s_\text{limit}$\;
                $A^\ast = \operatorname{copy}(A)$\;
                reconnect edges in $A^\ast$ according to $s^\ast$\;
                add $(L, r(A^\ast))$ to $\Gamma$\;
                $\Delta r^\ast = r_\text{target} - r(A^\ast)$\;
            }
        }
        $A = A^\ast$\;
        $\Delta r = \Delta r^\ast$\;
    }
\end{algorithm}

\subsection{Permutation Method}

The well-known configuration model for generating adjacency matrices~\cite{new03} typically includes auto-connections and multiple edge connections with no control over their appearance or proportion, often forcing post-processing removal if none are desired. We developed a novel adjacency matrix assembly technique that, given predefined sequences of in- and out-degrees, permits designating not only whether multiple edges appear but also their proportion --- with no post-processing required. Additionally, auto-connections can be included or omitted without manipulating the resulting $A$. These $A$'s exhibit generally neutral assortativities over all types with exceptions emerging for the highest multi-edge densities we assembled: e.g., 98-99\% multi-edges exceed our target neutral assortativity range of $\pm 0.005$, so for purposes of this study these were discarded.

\begin{figure}
    \includegraphics[width=.8\linewidth]{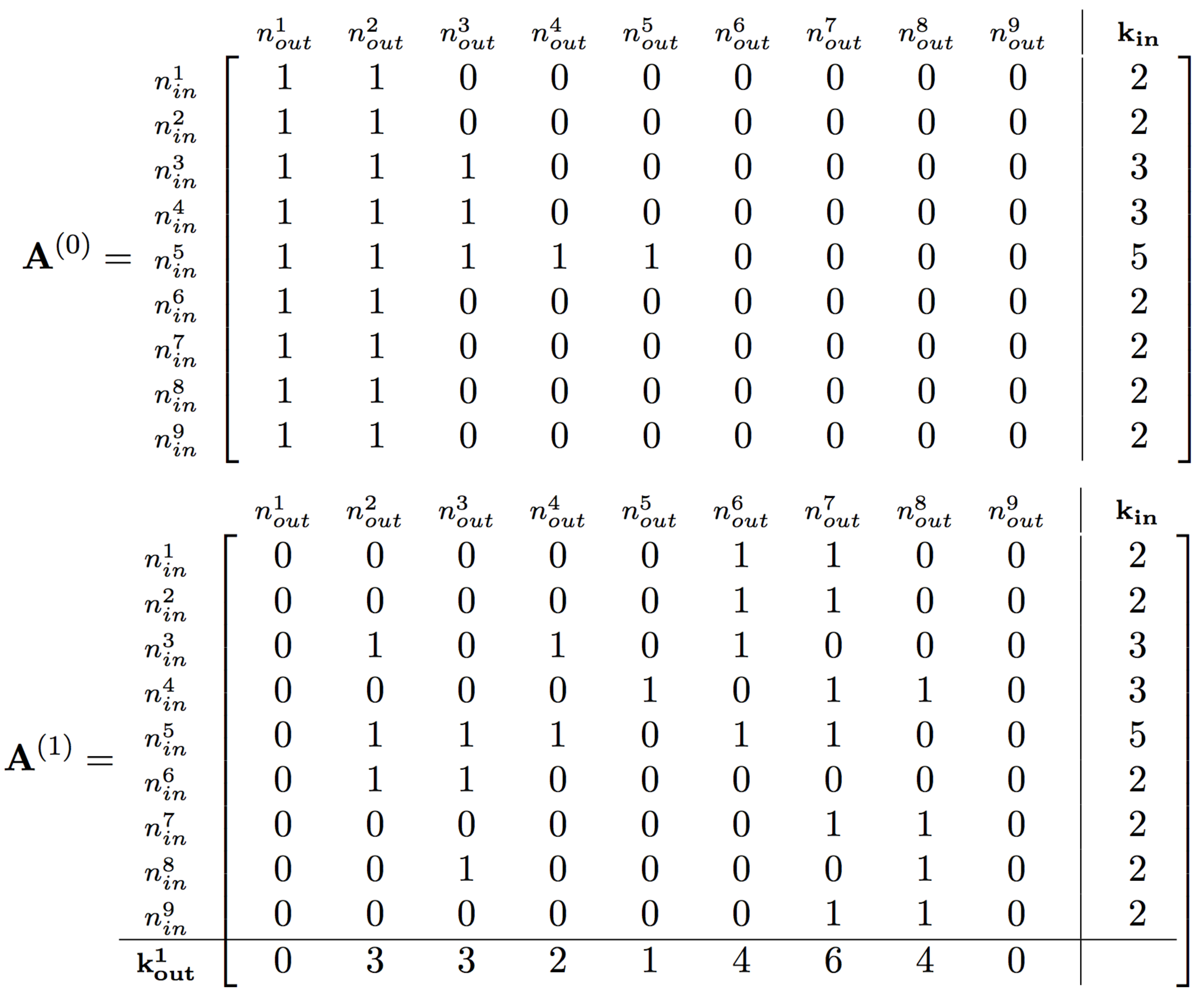}
    \caption{Permutation method initial matrices illustration. Top: $A^{(0)}$ showing arrangement of edge entries (all solo connections: $\rho_{m}^{+}=0$) for each row aligned left where row sums add up to $k_{in}$. If multi-edges are desired, we simply distribute them in the rows of $A^{(0)}$ satisfying the proportion, $\rho_{m}^{+}=0$ and the row sum. Bottom: permutation of rows in $A^{(0)}$ into this example $A^{(1)}$. Note row sums still add up to $k_{in}$, with column sums adding to a current $k_{out}^{(1)}$ --- likely violating the designated $k_{out}$. }
    \label{fig:permute_eg_A}
\end{figure}

This permutation method is a two-phase approach requiring only two sequences of in- and out-degrees, or $k_{in}$ and $k_{out}$, respectively, and a target multiple edge connection density, $\rho_{m}^{+}$. We describe this technique briefly here, and with more detail in a subsequent companion publication. These phases are as follows:

\begin{enumerate}
	\item Generate an initial matrix, designated $A^{(0)}$, with each node's inbound edge counts (row sums) satisfying $k_{in}$, yet ignoring $k_{out}$.  Each row of $A^{(0)}$ is filled with nonzero entries comprised of solo- and multiple-edge connections whose sum is $k_{in}$ for each node. Remaining entries along each row are simply filled with zeros out to the $N^{th}$ column. This resulting $A^{(0)}$ thus adheres to the designated $k_{in}$ and $\rho_{m}^{+}$, but violates $k_{out}$: all the column sums are incorrect (see Fig. (\ref{fig:permute_eg_A})).
	\item Randomly permute each row of $A^{(0)}$, distributing the non-zero entries of solo- and multiple-edge connections into a first sequence of permuted matrices, $A^{(1)}$. Calculate an error distance for $A^{(1)}$ from the designated $k_{out}$ via $e_{out}^{(1)} = k_{out} - k_{out}^{(1)}$. Each entry in $e_{out}^{(1)}$ is used to classify nodes: too many out-bound edges (``donor'' nodes), those with too few (``recipient'' nodes) and those at their target out-degree (``inert'' nodes). We then loop over all the donor nodes, randomly pick a recipient node and exchange edges from the donor to the recipient --- if suitable. After each exchange, we update the current permuted matrix, $A^{(i)}$, its corresponding error distance, $e_{out}^{(i)}$, and repeat the process until this error is zero. The final matrix, $A^{(n)}$, after $n$ updates, then satisfies all the designated characteristics if we performed suitable edge exchanges along the way.

\end{enumerate}

%\begin{align}
%\label{eqn:permute_A0}
%\mathbf{A}^{(0)} = 
%\kbordermatrix{
% & n^{1}_{out} & n^{2}_{out} & n^{3}_{out} & n^{4}_{out} & n^{5}_{out} & n^{6}_{out} & n^{7}_{out} & n^{8}_{out} & n^{9}_{out} & \vrule & \mathbf{k_{in}} \\
%n^{1}_{in} & 1 & 1 & 0 & 0 & 0 & 0 & 0 & 0 & 0 & \vrule & 2 \\
%n^{2}_{in} & 1 & 1 & 0 & 0 & 0 & 0 & 0 & 0 & 0 & \vrule & 2 \\
%n^{3}_{in} & 1 & 1 & 1 & 0 & 0 & 0 & 0 & 0 & 0 & \vrule & 3 \\
%n^{4}_{in} & 1 & 1 & 1 & 0 & 0 & 0 & 0 & 0 & 0 & \vrule & 3 \\
%n^{5}_{in} & 1 & 1 & 1 & 1 & 1 & 0 & 0 & 0 & 0 & \vrule & 5 \\
%n^{6}_{in} & 1 & 1 & 0 & 0 & 0 & 0 & 0 & 0 & 0 & \vrule & 2 \\
%n^{7}_{in} & 1 & 1 & 0 & 0 & 0 & 0 & 0 & 0 & 0 & \vrule & 2 \\
%n^{8}_{in} & 1 & 1 & 0 & 0 & 0 & 0 & 0 & 0 & 0 & \vrule & 2 \\
%n^{9}_{in} & 1 & 1 & 0 & 0 & 0 & 0 & 0 & 0 & 0 & \vrule & 2 \\ 
%%\mathbf{k_{out}} & 1 & 4 & 1 & 1 & 1 & 2 & 3 & 9 & 1 & \vrule \\
%}
%\end{align}

\begin{algorithm}
    \caption{Permutation Method, Phase (1). \newline
    \textit{Pseudo-code for generating initial permutation matrix $A^{(0)}$ satisfying $k_{in}$ and $\rho_{m}^{+}$ yet violating $k_{out}$. }}
%    \SetAlgoLined
    \label{alg:permute_phase1}

    \tcc{compute number of multi-edges per node}
    $k_{m}^{+} = \rho_{m}^{+}*k_{in}$ \;
    \For{\upshape $i=1...N$}{
    	\tcc{assemble vector possible edge values}
	$\mathbf{e_i} = [1,...,max(k_{m}^{+}(i))]$ \;
	\tcc{pick random edge values until sum of picks = $k_{in}(i)$}
	\While{\upshape sum($\mathbf{p}$) $\ne k_{in}$}{
		$p_t$ = \texttt{randpick}($\mathbf{e_i}$) \;
		$k_{temp}$ = sum($[\mathbf{p}$  $p_t$]) \;
		\eIf{\upshape $k_{temp} \le k_{in}(i)$}{
			append $p_t$ to $\mathbf{p}$ \;
		}{
			reject $p_t$ \;
		}	
	}
	\tcc{update $A^{(0)}(i,:)$ with row vector $\mathbf{p}$ and zeros out to $N$}
	$A^{(0)}(i,:) = [\mathbf{p}$ ... 0$]_{(1,N)}$ \;
    }
\end{algorithm}

\begin{algorithm}
    \caption{Permutation Method, Phase (2). \newline
    \textit{Pseudo-code for permuting initial matrix $A^{(0)}$ into final matrix $A^{(n)}$ satisfying all constraints, $k_{in}$, $\rho_{m}^{+}$ and $k_{out}$. }}
%    \SetAlgoLined
    \label{alg:permute_phase2}

\tcc{permute entries of $A^{(0)}$ into $A^{(1)}$ }
\For{\upshape $i=1...N$}{
	$A^{(1)}(i,:) = \texttt{randperm}(A^{(0)}(i,:))$ \;
}    
\tcc{compute deviation from target $k_{out}$}
$\mathbf{e}_{out}^{(1)} = k_{out}^{(1)} - k_{out}$ \;
$e_{norm}^{(1)}= \texttt{norm}(\mathbf{e}_{out}^{(1)})_{L2}$ \;
\tcc{classify node deviation type}
$\mathbf{d}^{(1)} = \mathbf{e}_{out}^{(1)} > 0$ \;
$\mathbf{r}^{(1)} = \mathbf{e}_{out}^{(1)} < 0$ \;
\tcc{loop over donors until target $k_{out}$ reached}
\While{\upshape $e_{norm}^{(i)} \ne 0$}{
    \For{\upshape $j$ \texttt{over} $\mathbf{d}^{(i)}$}{	
	\tcc{pick suitable pair donor and recipient entries}
    	\While{\text{swap pair unsuitable}}{
        		$k = \texttt{randpick}(A^{(i)}(:,j))$ \;
        		%$l = \texttt{randpick}(\mathbf{r}^{(i)})$ \;
		\tcc{if autoconnects disabled, remove this column from recipient list}
		\If{no autoconnects}{
			$\mathbf{r}^{(i)} = \texttt{remove}(\mathbf{r}^{(i)},k)$ \;
		}
		 $l = \texttt{randpick}(\mathbf{r}^{(i)})$ \;
		\If{$A^{(i)}(k,j) \geq A^{(i)}(k,l)$}{
			\emph{swap pair suitable} \;
		
		}

	}
	\tcc{exchange edge from donor to recipient}
	$A^{(i)}(k,j) = A^{(i)}(k,j) - 1$ \;
	$A^{(i)}(k,l) = A^{(i)}(k,l) + 1$ \;
	\tcc{update error tracking}
	$\mathbf{e}_{out}^{(i)} = k_{out}^{(i)} - k_{out}$ \;
	$e_{norm}^{(i)}= \texttt{norm}(\mathbf{e}_{out}^{(i)})_{L2}$ \;
	\tcc{update recipient list}
	$\mathbf{r}^{(i)} = \mathbf{e}_{out}^{(i)} < 0$ \;

     }
     \tcc{update donor list}
	$\mathbf{d}^{(i)} = \mathbf{e}_{out}^{(i)} > 0$ \;

}
    
\end{algorithm}

%\bibliographystyle{plain}
%\bibliography{chim2}

\end{document}